\documentstyle{article}

\def\etal{{et al.\,\,}}

\def\kms{${\rm km\,s^{-1}}$}

\def\Hone{H{\sc i}}

\def\bootes{Bo\"{o}tes}
\def\iras{{\it IRAS}}

\begin{document}
\title{An \Hone\ survey of the Bo\"{o}tes void.  II.  The Analysis}
\author{Arpad Szomoru \thanks{Kapteyn Astronomical Institute, P.O.  Box
800, NL 9700 AV Groningen, The Netherlands, Visiting Astronomer, Kitt
Peak National Observatory, National Optical Astronomy Observatory} \and
J.  H.  van Gorkom \thanks{Columbia University, 538 W 120${\rm th}$
Street, New York NY 10027, USA} \and Michael D.  Gregg \thanks{Institute
of Geophysics and Planetary Physics, Lawrence Livermore National
Laboratory, P.O.  Box 808, L-413 Livermore, CA 94551-9900, USA, Visiting
Astronomer, Kitt Peak National Observatory, National Optical Astronomy
Observatory} \and Michael A.  Strauss \thanks{Princeton University Obs.,
Peyton Hall, Princeton, NJ 08544, USA}}
\date{ }

\maketitle

\begin{abstract} We discuss the results of a VLA \footnote{The NRAO is
operated by Associated Universities, Inc., under a cooperative agreement
with the National Science Foundation.} (Napier \etal 1983) \Hone\ survey
of the \bootes\ void and compare the distribution and \Hone\ properties
of the void galaxies to those of galaxies found in a survey of regions
of mean cosmic density.  The \bootes\ survey covers 1100 Mpc$^{3}$, or
$\sim$ 1\% of the volume of the void and consists of 24 cubes of
typically 2 Mpc $\times$ 2 Mpc $\times$ 1280 \kms, centered on optically
known galaxies.  Sixteen targets were detected in \Hone; 18 previously
uncataloged objects were discovered directly in \Hone.  The control
sample consists of 12 cubes centered on \iras\ selected galaxies with
FIR luminosities similar to those of the \bootes\ targets and located in
regions of 1 to 2 times the cosmic mean density.  In addition to the 12
targets 29 companions were detected in \Hone.  We find that the number
of galaxies within 1 Mpc of the targets is the same to within a factor
of two for void and control samples, and thus that the small scale
clustering of galaxies is the same in regions that differ by a factor of
$\sim$ 6 in density on larger scales.

A dynamical analysis of the galaxies in the void suggests that on scales
of a few Mpc the galaxies are gravitationally bound, forming interacting
galaxy pairs, loose pairs and loose groups.  One group is compact enough
to qualify as a Hickson compact group (hereafter HCG) (Hickson 1982).

The galaxies found in the void are mostly late-type, gas rich systems.
A careful scrutiny of their \Hone\ and optical properties shows them to
be very similar to field galaxies of the same morphological type.  This,
combined with our finding that the small scale clustering of the
galaxies in the void is the same as in the field, suggests that it is
the near environment that mostly affects the evolution of galaxies.

\end{abstract}

\end{document}